\documentclass[aps,prl,twocolumn,groupedaddress,amsmath,amssymb,superscriptaddress]{revtex4}
\usepackage{xfrac}
\usepackage{amsmath,amssymb,bm}
\usepackage{graphics}
\usepackage{epsfig}
\usepackage{graphicx}
\usepackage{adjustbox}
\usepackage{caption}
\usepackage{natbib}
\usepackage{rotating}
\usepackage{multirow}
\usepackage{upgreek}
\usepackage{color}
\usepackage{enumerate}

\newcommand{\mean}[1]{\left\langle#1\right\rangle}

\usepackage{xcolor}

\usepackage{setspace}
\usepackage{siunitx}
\usepackage{rotating}

\usepackage{url}
  \let\oldurl\url
\usepackage{hyperref}
  \let\linkurl\url
  \let\url\oldurl

\begin{document}
\makeatletter
\newcommand{\thickhline}{%
    \noalign {\ifnum 0=`}\fi \hrule height 1.5pt
    \futurelet \reserved@a \@xhline
}
\newcolumntype{"}{@{\hskip\tabcolsep\vrule width 1.5pt\hskip\tabcolsep}}
\makeatother
\begin{abstract}
The concept of the cosmic web, viewing the Universe as a set of discrete galaxies held together by gravity, is deeply engrained in cosmology. Yet, little is known about the most effective construction and the characteristics of the underlying network.  Here we explore seven network construction algorithms that use various galaxy properties, from their location, to their size and relative velocity, to assign a network to galaxy distributions provided by both simulations and observations.  We find that a model relying only on spatial proximity offers the best correlations between the physical characteristics of the connected galaxies. We show that the properties of the networks generated from simulations and observations are identical, unveiling a deep universality of the cosmic web. 
\end{abstract}
\title{The Network Behind the Cosmic Web}
\author{B.~C. Coutinho}
\affiliation{Center for Complex Network Research and Department of Physics, Northeastern University, Boston, Massachusetts 02115, USA}

\author{Sungryong Hong}
\affiliation{National Optical Astronomy Observatory, 950 N. Cherry Ave., Tucson, AZ 85719}
\affiliation{The University of Texas at Austin, Austin, TX 78712, USA}

\author{Kim Albrecht}
\affiliation{Center for Complex Network Research and Department of Physics, Northeastern University, Boston, Massachusetts 02115, USA}

\author{ Arjun Dey}
\affiliation{National Optical Astronomy Observatory, 950 N. Cherry Ave., Tucson, AZ 85719}

\author{Albert-L\'aszl\'o Barab\'asi}
\affiliation{Center for Complex Network Research and Department of Physics, Northeastern University, Boston, Massachusetts 02115, USA}
\affiliation{Center for Cancer Systems Biology, Dana-Farber Cancer Institute, Boston, Massachusetts 02115, USA}
\affiliation{Department of Medicine and Channing Division of Network Medicine, Brigham and Women’s Hospital, Harvard Medical School, Boston, Massachusetts 02115, USA}
\affiliation{Center for Network Science, Central European University, 1051, Budapest, Hungary}

\author{Paul Torrey}
\affiliation{MIT Kavli Institute for Astrophysics and Space Research, 77 Massachusetts Ave. 37-241, Cambridge MA 02139, USA}
\affiliation{California Institute of Technology, Pasadena, CA 911, USA}

\author{Mark Vogelsberger}
\affiliation{Harvard-Smithsonian Center for Astrophysics, 60 Garden Street, Cambridge, MA 02138, USA}

\author{Lars Hernquist}
\affiliation{Harvard-Smithsonian Center for Astrophysics, 60 Garden Street, Cambridge, MA 02138, USA}
\maketitle

The cosmic web, the desire to view the large-scale structure of the Universe as a network, is deeply embedded both in cosmology and in public consciousness \cite{citeulike:13161782,PhysRevD.81.103006,citeulike:13346231,cite-key,Faucher}. Yet, it remains little more than a metaphor, typically used to capture the dark matter's  ability to agglomerate the galaxies in a web-like-fashion. Numerous halo finder algorithms \cite{citeulike:13405789,citeulike:9111747}, made possible by the increasingly precise simulations of the evolution of the Universe \cite{citeulike:13161782,2005Natur.435..629S}, exploit the  network-like binding of the galaxies \cite{citeulike:8909637}. Yet, very little is known about the graph theoretical characteristics of the resulting cosmic web.  Our goal here is to test and explore various meaningful definitions of the cosmic web, and use the tools of network science to characterize the generated networks. In particularly, we explore which network definition offers the best description of the observed correlations between the physical characteristics of  connected galaxies. The resulting network-based framework, tested in both simulations and observational data, offers a new tool to investigate the topological properties of the large scale structure distribution of the Universe.  

We start with data provided by a subhalo catalog constructed from the Illustris~\cite{citeulike:13161782,Vogelsberger21102014,Nelson201512} cosmological simulation that traces the growth of large scale structure, galaxy formation and evolution from 2Gy after the Big Bang to the present epoch, incorporating both baryons and dark matter. In line with common practice, we assume that subhalos in the simulation correspond to galaxies in the observational data \cite{0004-637X-683-1-123}, representing the \textit{nodes} of the cosmic web. By considering all subhalos with stellar mass bigger than  $M_{*}>10^9~M_{\text{sun}}$, we obtain between 2,000 and 30,000 subhalos for different redshifts (supplementary material~A).

\begin{figure*}[!t]
\captionsetup{justification=raggedright,
singlelinecheck=false
}
\begin{center}
 \includegraphics[width=1\textwidth]{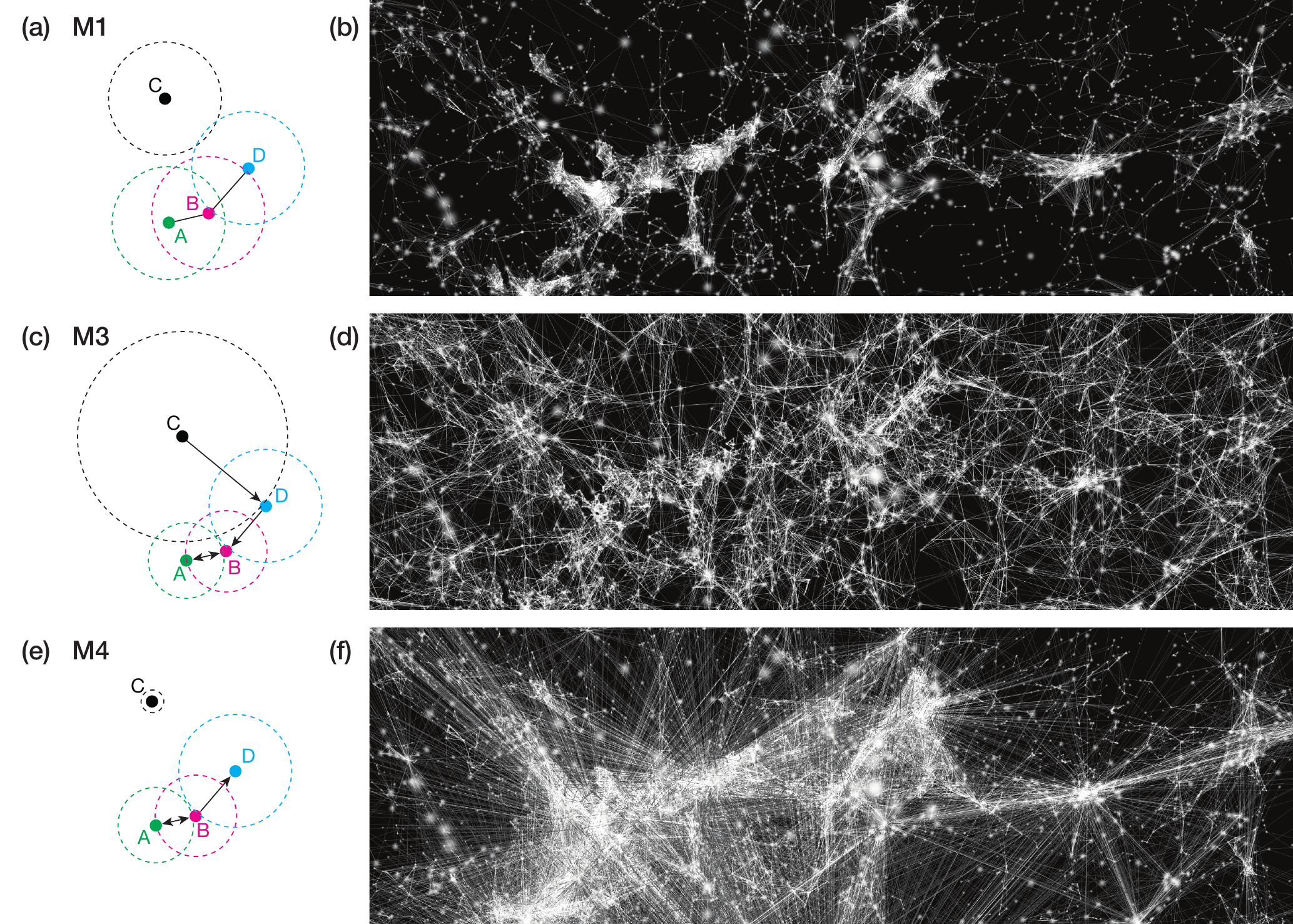}
\end{center}
\caption{\textbf{Building networks from galaxy data}. The circles represent the linking lengths for models M1, M3 and M4. (a) In M1 all galaxies within distance $l$ are connected by an undirected link. (c) In M3 a galaxy is connected to the closest galaxy with a directed link; therefore the  linking length depends on the position of the closest galaxy. (e) In M4, the linking length scales with the galaxy size, $l=a~R_i^{1/2}$. (b),(d) and (f) Visualization of the cosmic web for redhsiht $0$ produced by the respective models, for $\mean{k}=40$. For simplicity the direction of the links is not present in the visualization. For interactive visualization see \linkurl{http://kimalbrecht.com/ccnr/04-networkuniverse/17-network-interface}. Models M2,5,6,7 are generated from the three models shown above. In M2 the directions of the M3 links are inverted; in M5 the direction of the M4 links are inverted. M6(7) are similar to M4(5) but computed in the phase space.\label{Sequematic}}
\end{figure*}

There are multiple ways of building networks from the available subhalo/galaxy catalogs, allowing us to define seven distinct models for the construction of the cosmic web (M1-M7). The simplest, M1, links two nodes with an undirected link if the distance between them is smaller than a predefined length, $l$ (Fig.\ref{Sequematic}(a) and (d)). M2(3) represent the directed versions of M1, drawing a directed link $j\rightarrow i$ ($i\rightarrow j$) from $i$ to the closest $\mean{k}$ nodes (Fig.\ref{Sequematic}(b) and (e)). Consequently, while in  M1 a node $i$ can have arbitrary degree (number of neighbors connected to $i$), in M2(3) the in(out) degrees are fixed (number of neighbors connected to $i$ through a directed link $j\rightarrow i$($i\rightarrow j$)). In M4(5) a directed link $j\rightarrow i$ ($i\rightarrow j$) is drawn from $i$ to $j$ if the distance between the two nodes is smaller than $a\cdot R_i^{1/2}$, where $a$ is a free parameter and $R_i^{1/2}$ is the half-mass radius~\cite{2008gady.book.....B}. Models 6(7) are extensions of  M4(5), but computed in phase space, where a directed link $j\rightarrow i$ ($i\rightarrow j$) is drawn from $i$ to $j$ if the sum of the square of the normalized distance and relative speed between two nodes is smaller than $a^2$. Taken together, M1-3 require only data about the halo/galaxy positions,  M4(5) require information about galaxy positions and size and M6(7) require galaxy velocities and linking galaxies that may be gravitationally bound (supplementary material~B offers the formal definition of each model).

The distinct network representations of the cosmic web, offered by the models introduced above, raises the question: Which of these representations are the most meaningful? In general, networks are only meaningful if the links have functional roles, linking either interacting nodes or nodes with similar characteristics. For example, the links of a social network tend to connect individuals with similar social-economic characteristics (homophily) and in cellular networks connected proteins tend to have related biological roles.  The fact that the color of a satellite galaxy is correlated with the mass of the host galaxy~\cite{citeulike:13411806,2011MNRAS.417.1374P,2012MNRAS.424..232W,2013MNRAS.434.1838G,2014MNRAS.442.1363W,Bray01012016} indicates that such correlations between nearby galaxies are meaningful. We therefore explore the degree to which the above network representations of the cosmic web add links between galaxies/subhalos of similar physical characteristics. For this we analyze 71 parameters that characterize each subhalo, ranging from their peculiar velocity  to star formation rate (supplementary material~C for the entire list), allowing us to identify the network representation that offers the best correlation between them. Since we are working from a cosmological simulation, some of the correlations may be meaningful only in the sense that they characterize the underlying properties (or assumptions) of the model. Nevertheless, our analyses provides an unbiased way of probing the spatial network without any a priori biases. 

For a given model M and subhalo property $c_i$,  we compute the average value of $c_i$ over all nodes connected to $i$, 
\begin{equation}
\tilde{c}_i\equiv \frac{\sum_j a_{ij} c_j}{k_i},
\end{equation}
where $k_i$ is the degree of node $i$ and $a_{ij}$ is the adjacency matrix. We use the Pearson coefficient to  measure correlations between the connected nodes, 
\begin{equation}
 R\equiv\frac{\sum_{i}(\tilde{c}_i-\mean{\tilde{c}})(c_i-\mean{c})}{\sqrt{\sum_{i}(\tilde{c}_i-\mean{\tilde{c}})^2\sum_{i}(c_i-\mean{c})^2}},\displaystyle
 \end{equation}
where $\mean{c}$ and $\mean{\tilde{c}}$ are the average of $c_i$ and $\tilde{c}_i$ over all nodes. Since the scale over which correlations persist is unknown, we construct networks with different average degrees, $\mean{k}$. 

\begin{figure}[!t]
\centering
\captionsetup{justification=raggedright,
singlelinecheck=false
}
 \includegraphics[width=0.5\textwidth]{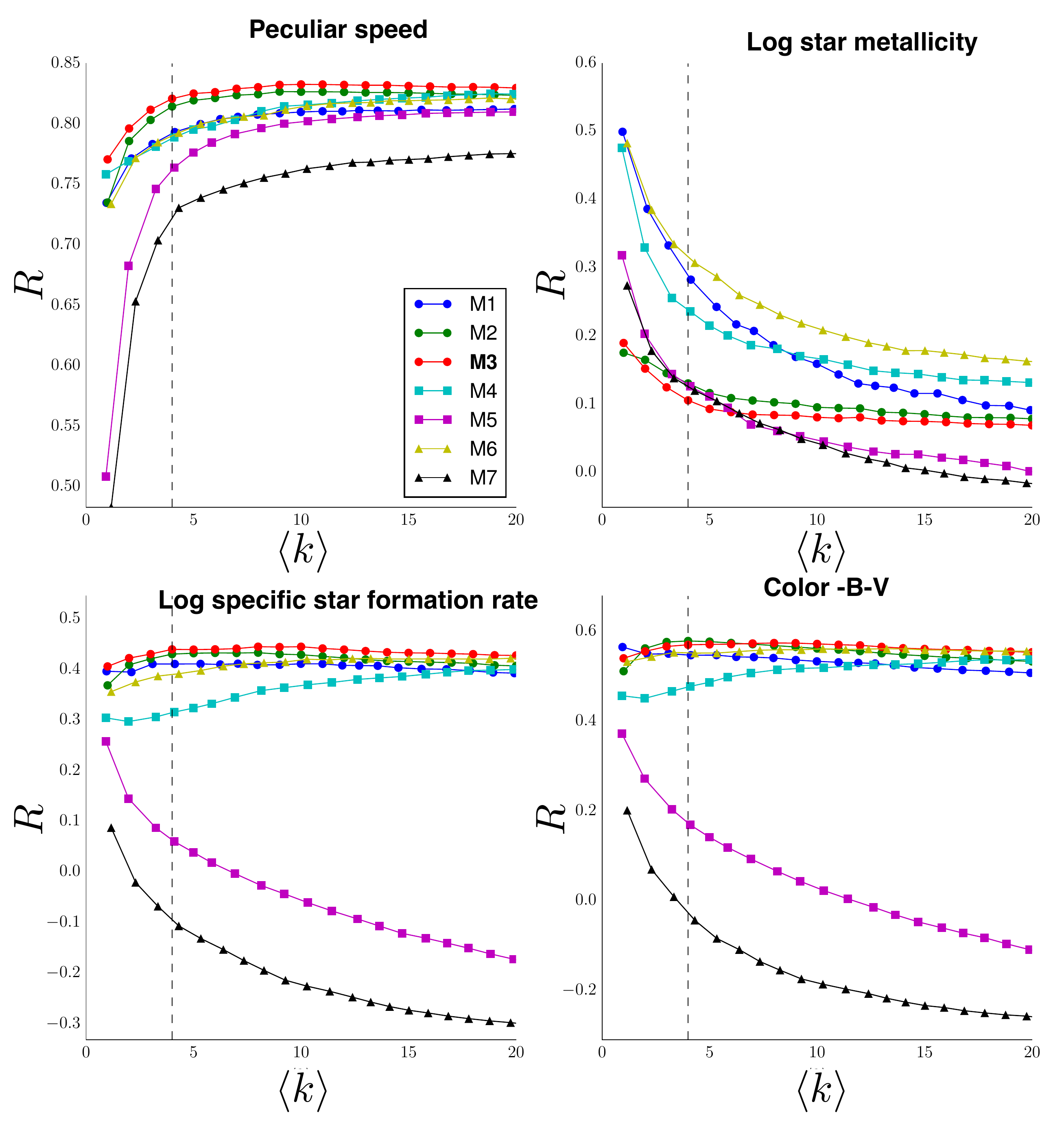}
\caption{\textbf{Correlations between connected galaxies.}The Pearson coefficient $R$ capturing the correlation between connected nodes as a function of the mean degree $\mean{k}$ for all algorithms, redhsiht $0$ and for various galaxy properties. The giant strongly connected component emerges at $\mean{k}=4$ for M3, shown as a dashed vertical line.\label{R_vs_degree_all_recipes}}
\end{figure}

We find four properties that  consistently display correlations between the connected nodes: peculiar velocity, stellar metallicity, specific star formation rate, and color in the B-V band (Fig.~\ref{R_vs_degree_all_recipes}). We also find that of the seven models, M3 captures the best correlations for the peculiar speed, specific star formation rate and rest-frame B-V color. It only  fails to maximize the correlation between stellar metallicities, in which M6 excels. We also calculated the correlation function for networks obtained under node and link randomization (supplementary material B). The lack of significant correlations in these randomized networks indicates that Fig.~2 captures physical meaningful correlations between connected galaxies. Note that some of these correlations, like the metallicity, may result from the assumptions made by the simulation, while others, like the peculiar velocity, likely reflect physical correlations in the real Universe. However, this exercise demonstrates that it is indeed possible to uncover underlying properties of the network without prior knowledge. Overall, Fig.~2 indicates that model M3 captures best the correlations between the properties of connected galaxies. Its superiority over the more data demanding models M4-M7 suggests that spatial proximity, despite its simplicity, remains the most powerful organizing principle of the cosmic web. 
Given the ability of M3 to best capture correlations between the subhalo characteristics, in the remainder of the paper we analyze the networks predicted by this model.

\begin{figure}[!t]
\centering
\captionsetup{justification=raggedright,
singlelinecheck=false
}
 \includegraphics[width=0.5\textwidth]{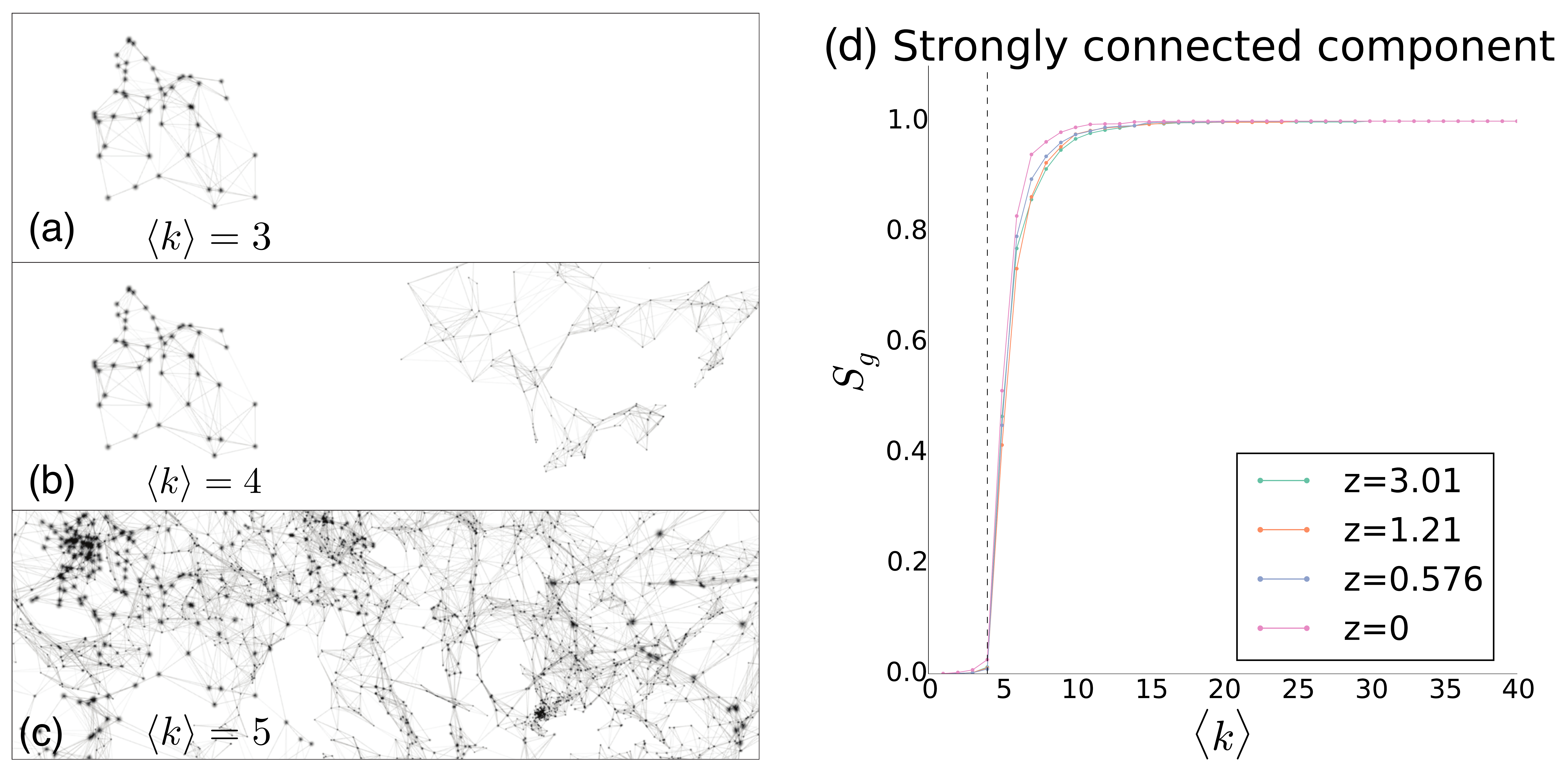}
\caption{\textbf{The evolution of the cosmic web.} (a), (b) and (c), visualization of the largest strongly connected component for different $\mean{k}$. Links that cross the boundaries are ignored in the plots, but not in the computation of the components. (d) The largest strongly connected component, $S_g$, as a function of the mean degree, $\mean{k}$, for redshifts $0\leq z\leq 3.01$. The giant strongly component emerges at $\mean{k}=4$, marked by a dashed line. \label{wlgc_c_1_2_3}}
\end{figure}

The overall integrity of a network is well characterized by the size of its largest connected component~\cite{Barabasi_book}. For a directed network the \textit{strongly connected component} is the largest subset of nodes such that for all pairs  $i$ and $j$ in the subset there is a directed path from $i$ to $j$. Figure \ref{wlgc_c_1_2_3} illustrates that the giant strongly connected component emerges at $\mean{k}=4$ for all redshifts in M3. At the practical level, this implies that M3 can be applied at different redshifts and number of nodes without the need to adjust the model parameters. At a more fundamental level, it means that the critical mean degree of the giant strongly connected component is universal, being rooted in the intrinsic proprieties of the galaxy distribution.

To further validate M3, we compare the structure of the cosmic web obtained in the simulations with observational data from the Sloan Digital Sky Survey (SDSS)~\cite{citeulike:4646910,citeulike:3910363}, that provides information about the position and the properties of the galaxies in the visible sky \cite{galaxy_populations}. We study the section of the sky with redshift $z<0.03$, right ascension $100<RA<270$ and declination $-7<DEC<70$. As a reference, we also study a randomized version of the data by distributing the galaxies  randomly in space (Appendix G).
To confirm that the galaxy population in the simulation and in the observations are comparable in the same redshift range, we measured the galaxy mass function, defined as the number of galaxies with stellar mass $M_{*}$ per volume. Since the density of galaxies with stellar mass under $10^{9} M_{sun}$ is much higher in the simulation than in the  observational data, a known limitation of the simulation \cite{galaxy_populations}, we only consider galaxies with stellar mass $M_{*}>10^{9} M_{sun}$.
 \begin{figure}[!t]
\centering
\captionsetup{justification=raggedright,
singlelinecheck=false
}
 \includegraphics[width=0.5\textwidth]{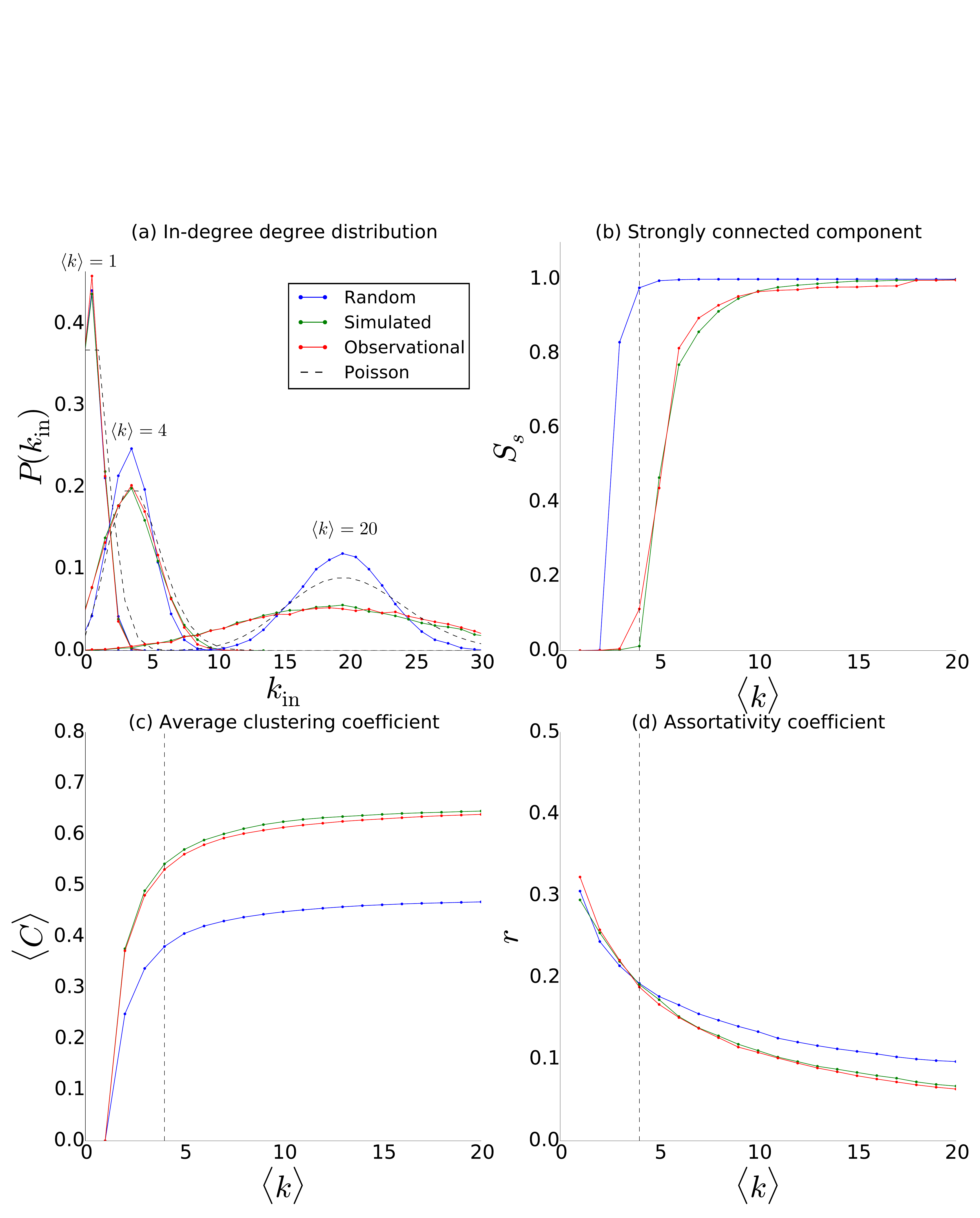}
\caption{\textbf{Network characteristics of the cosmic web.} (a) In-degree distribution for mean degrees $\mean{k}=1,4$ and $20$. The dashed lines represents the Poisson distribution. (b) Size of the largest strongly connected component, $S_g$, as a function of the mean degree, $\mean{k}$. (c)  Average clustering coefficient as a function of the average mean degree $\mean{k}$. (d) Assortativity coefficient, $r$, as a function of $\mean{k}$. Each panel show data for M3, for the random, the simulated (redshift $0$) and the observational networks.
\label{c_a_w_s}}
\end{figure}

Figure~\ref{c_a_w_s}(a) documents an excellent agreement between the in-degree distribution (fraction of nodes with a given in-degree), $P(k_{\mathrm{in}})$, for the observational and the simulated M3 networks. Both distribution deviate from the random distribution, indicating that the observed $P(k_{\mathrm{in}})$ reflects the non-trivial galaxy distribution in both the simulations and in the observations. We show analytically in supplementary material E, that networks constructed by the M3 model for a random galaxy distribution, the variance of the in-degree distribution  is $0.709$ for $\mean{k}=1$. The fact that the variance is smaller than 1, which is the value expected for a Poisson distribution, implies that the degree distribution is narrower than that expected for an Erd\H{o}s{-}R\'enyi network, hence hubs are definitely absent, quite distinct from what is found in biological and social networks, were hubs are prevalent.
We also obtain excellent agreement between the simulation and observation based networks for the average clustering coefficient, capturing the fraction of triangles in the network, and assortativity (Figure \ref{c_a_w_s}(c) and (d)). In both cases the simulation and observation-based values agree with each other, both deviating from the random expectation. The giant strongly connected component emerges at $\mean{k}=4$ for both the simulation and observational M3 networks, while it is at $\mean{k}=3$ for the random M3 networks (Figure \ref{c_a_w_s} (b)). The results indicate that M3 offers an accurate description of the cosmic web, capturing consistently its network characteristics,  both in the simulation and in the observational data.

In summary, here we used the tools of network science to characterize the large structure of the Universe both in simulations and observational data. While we can define numerous network construction algorithms, we find that the simple model M3, which relies on spatial proximity only, captures the best correlations between the physical characteristics of nearby galaxies. The results are distinct from the random case, which assumes random galaxy localizations, indicating that the obtained structure of the cosmic web is intricately tied to the underlying structure of the Universe. It is particular encouraging that the network characteristics of the cosmic web, from the degree distribution to the clustering and degree correlations, show remarkable agreement between simulations and observations.    In many ways, our results represent only the first step towards a network-based understanding of the Universe. Yet, they provide guidance for the nature of the data needed for a systematic exploration of the underlying network, offering a framework on with one could build various applications, from halo finders to exploring the fundamental characteristics of the cosmic web.

This research was supported in part by the National Optical Astronomy Observatory (NOAO) and by the Radcliffe Institute for Advanced Study and the Institute for Theory and Computation at Harvard University. NOAO is operated by the Association of Universities for Research in Astronomy (AURA), Inc. under a cooperative agreement with the National Science Foundation, and by the FCT Grant No. SFRH/BD/79723/2011.

 \newpage

\clearpage
 \bibliography{bib}

\end{document}